\documentstyle[11pt]{article}
\begin{document}
\begin{titlepage}
\begin{center}

\vspace{2cm}

\Large \bf{Integrable models of interacting quantum
spins with competing interactions}

\vspace{1cm}
\end{center}
\begin{flushleft}
\normalsize {Johannes Richter (*), Sven E. Kr\"uger(*),
Andreas Voigt (*) and Claudius Gros (**)}\\
\vspace{0.3cm}
\it {(*)Institut f\"ur Theoretische Physik, Universit\"at Magdeburg\\
P.O.B. 4120, D-39016 Magdeburg, Germany\\
(**)Institut f\"ur Physik, Universit\"at Dortmund\\
D-44221 Dortmund, Germany\\}
\end{flushleft}

\begin{abstract}
We present a class of exactly solvable quantum spin models which consist
of two Heisenberg-subsystems coupled via a long-range Lieb-Mattis
interaction. The total system is exactly solvable whenever the
individual subsystems are solvable and allows to study the effects of
frustration. We consider (i) the antiferromagnetic linear chain and (ii)
the Lieb-Mattis antiferromagnet for the subsystem-Hamiltonians and
present (i) the complete ground-state phase diagram and (ii) the full
thermodynamic phase diagram. We find a novel phase which exhibits {\it
order from disorder} phenomena. \end{abstract}


\end{titlepage}



{\it Introduction.} --
The properties of interacting quantum spins have been studied
intensively over a long period. Some selected model Hamiltonians, which
typically describe homogeneous collections of interacting spins, can be
solved exactly. Examples are (i) models in one dimension solvable by
Bethe-Ansatz \cite{bethe}, (ii) a one-dimensional model with long-range
inverse-square exchange \cite{haldane}, (iii) valence-bond models
\cite{majum} and (iv) models with long-range interaction of constant
magnitude (Lieb-Mattis type models \cite{lieb62} which have recently
been  used to discuss spontaneous symmetry breaking in spin systems
\cite{kaipe}). Our understanding of interacting quantum spins has
benefited greatly from these studies.

On the other hand, in experimental applications we have to deal
sometimes with situations where two subsystems of spins interact
mutually. One example is the garnet, Mn$_3$Cr$_2$Ge$_3$O$_{12}$, in
which the Mn and Cr moments are located on two disjunct,
inter-penetrating sublattices \cite{golos76,valyan78}, with the
intra-sublattice couplings dominating over the inter-sublattice coupling
due to frustration effects \cite{valyan78,shender82}. Another example is
the two-layer structure in certain high-T$_c$ compounds, like
YBa$_2$Cu$_3$O$_{6}$ \cite{tranquada92}, where the competition between
the inter-layer and the intra-layer couplings might
lead to a spin-gap \cite{monien93}.

Here we present a class of exactly solvable models which consist of two
subsystems coupled via a long-range Lieb-Mattis \cite {lieb62}
interaction. The total system is exactly solvable whenever the solutions
of the individual subsystems are known and it allows to investigate the
effects of competing interactions. We consider two different types of
subsystems, the linear chain and the Lieb-Mattis model, and present the
ground-state (GS) phase diagrams. In addition we present for the case of two
coupled Lieb-Mattis subsystems the full thermodynamic phase diagram. We
find a novel phase which exhibits {\it order from disorder} phenomena
 \cite{villain,shender82}.

{\it Model.} --
We consider a system of $N=N_1+N_2$ interacting quantum spins ($s=1/2$)
made up of two mutually interacting subsets described by the Hamiltonian
\begin{eqnarray}\label{hamiltonian} H \ =\ \alpha_1 H_1\{{\bf s}_i\}
+ \alpha_2 H_2\{{\bf s}_l\} + \alpha H_{int}\{{\bf s}_i,{\bf s}_l\}
\end{eqnarray}
with $ \alpha , \alpha_1 , \alpha_2 > 0$.
Here $H_1\{{\bf s}_i\}$ ($i=1,\dots,N_1$) and
$H_2\{{\bf s}_l\}$ ($l=N_1+1,\dots,N$) contain the interactions within
the first and the second subset of $N_1$ and $N_2$ spins respectively.
The antiferromagnetic ($\alpha>0$) interaction between the two subsets
of spins is taken to be of the Lieb-Mattis form \cite{lieb62},
$H_{int}\{{\bf s}_i,{\bf s}_l\} = {\bf S}_1\cdot{\bf S}_2/(N_1 N_2)$
with ${\bf S}_1=\sum_{i=1}^{N_1}{{\bf s}_i}$ and \linebreak
${\bf S}_2=\sum_{l=N_1+1}^{N}{{\bf s}_l}$.
We assume $H_1$ and $H_2$ to be spin-rotationally invariant. Then the
operators ${\bf S}^2=({\bf S}_1+{\bf S}_2)^2$, ${\bf S}_1^2$, ${\bf
S}_2^2$, $H$ , $H_1$ and $H_2$ all commute with each other.
The eigenvalues of Eq.\ (\ref{hamiltonian}) are given by
$ E_{S,S_1,S_2} =
{[\alpha /(2 N_1 N_2)]}
 \left[S(S+1) - S_1(S_1+1) - S_2(S_2+1)
\right]
+ \alpha_1 E_1 + \alpha_2 E_2 $,
where $S_1$, $S_2$, $E_1$, $E_2$ and $S$ denote the quantum numbers of
${\bf S}_1^2$, ${\bf S}_2^2$, $H_1$, $H_2$ and ${\bf S}^2$ respectively
with $S\in[|S_2-S_1|,S_2 + S_1]$. We consider the thermodynamic limit,
$N_1,\, N_2 \rightarrow \infty$, and define  the normalized quantum
numbers $x_{\gamma}=2 S_{\gamma}/N_{\gamma}$ ($0 \le x_{\gamma} \le 1$,
$\gamma=1,2$). The GS energy which is
realized for $S=|S_2-S_1|$, can then be written as
$
E_0= - \alpha \hspace{4pt} x_1x_2/4 + \alpha_1 E_1(x_1)+ \alpha_2
E_2(x_2),
$
where $E_1(x_1)$ and $E_2(x_2)$ are the lowest eigenvalues of $H_1$ and
$H_2$ within the subspace of a given magnetisation $x_1$ and $x_2$. For
a large class of antiferromagnetic Hamiltonians $H_{\gamma}$ a level
ordering $E_{\gamma}(x)\ge E_{\gamma}(x^{\prime})$ is valid for
$x>x^{\prime}$ \cite{lieb62}. Then a competition arises in $E_0$
 between the term proportional to $\alpha$ which
tends to maximize $x_{\gamma}$ and the terms proportional to
$\alpha_{\gamma}$ which tend to minimize $x_{\gamma}$, giving rise to
frustration effects.

We consider two models for the Heisenberg Hamiltonians $H_1\{{\bf
s}_i\}$ and $H_2\{{\bf s}_l\}$ entering Eq.\ (\ref{hamiltonian}) and set
$\alpha=1$ for the rest of this paper. The two models are the
antiferromagnetic linear chain (LC),
$H_{\gamma}^{LC}=(1/N_{\gamma})\sum_{i}{{\bf s}_i{\bf s}_{i+1}}$, and
the Lieb-Mattis (LM) model, \linebreak $H_{\gamma}^{LM} =
(2/N_{\gamma})^2 {\bf S}_{\bf \gamma}^A{\bf S}_{\bf \gamma}^B$, with
${\bf S}^{A}_{\gamma} = \sum_{i\in A\in\gamma}{\bf s}_i$ and ${\bf
S}^{B}_{\gamma} = \sum_{i\in B\in\gamma}{\bf s}_i$ being the total-spin
operators of the $A,\ B$ sublattice of the respective subsystem
($\gamma=1,2$). The scalings with $N_{\gamma}$ are chosen such that (i)
all energies become intensive in the thermodynamic limit $N_1,\, N_2
\rightarrow \infty$ and (ii) that for the fully polarized state
($x_{\gamma}=1$) the energies are
$E_{\gamma}^{LC}(1)=E_{\gamma}^{LM}(1)=1/4$. Both models can be solved
exactly. For the LM model in addition to ${\bf S}_{\gamma}^2$ also the
respective sublattice spin-operators, $({\bf S}^{A,B}_{\gamma})^2$,
commute with the Hamiltonian and the corresponding (normalized) quantum
numbers, $x^{A,B}_{\gamma}= (4/N_{\gamma})S_{\gamma}^{A,B}$ ($0 \le
x^{A,B}_{\gamma} \le 1$) are separately conserved.

For the LM model, mean-field theory becomes exact and quantum
fluctuations vanish identically in the thermodynamic limit. The energy
$E_{\gamma}(x_{\gamma})$ can be calculated as \linebreak
$E_{\gamma}^{LM}=x_{\gamma}^2/2-\frac{1}{4}$. On the other side, quantum
fluctuations are maximal for the antiferromagnetic LC. We computed
$E_{\gamma}^{LC}(x_{\gamma})$ by numerically solving the Bethe-Ansatz
equations for finite magnetisation $x_{\gamma}$ \cite{bethe}.

{\it Ground-state phase diagram}. --
The knowledge of $E_{\gamma}(x_{\gamma})$ ($\gamma=1,2$) allows to
determine the GS phase diagram, which we present in Fig.1. The
phase diagram is symmetric under $\alpha_1 \leftrightarrow \alpha_2$ and
the GS has minimal total spin $S=|S_2 - S_1|$. For small
$\alpha_1$ and $\alpha_2$ the antiferromagnetic inter-subsystem
interaction dominates and the subsystems are saturated ferromagnets
($x_1=x_2=1$) with parallel sublattice magnetisations. For large
$\alpha_1$ and $\alpha_2$ the antiferromagnetic intra-subsystem
interaction dominates and two subsystems effectively decouple into their
individual singlets ($x_1=x_2=0$). A line of first-order phase
transitions separates the spin-singlet region from a region near the
axis where one of the subsystems is partially magnetized and the other
one is fully magnetized. The states with partial magnetisation are
connected to the fully polarized state by a second-order phase
transition.

In Fig.2 we have plotted the GS expectation values of the
inter-sub\-system correlation function, $\langle {\bf s}_i{\bf s}_l
\rangle_{1,2}$ ($i$ and $l$ are in the first and second subsystem,
respectively), and the intra-subsystem correlation function $\langle
{\bf s}_i{\bf s}_j \rangle_{1,1}$ of the first subsystem (for the LC
case $j=i+1$, for the LM case $i\in A$ and $j\in B$) as a function of
$\alpha_1$ for fixed $\alpha_2=0.075$. For small $\alpha_1$ saturated
ferromagnetic long-range order within the subsystems ($\langle {\bf
s}_i{\bf s}_j \rangle_{1,1}=+1/4$ but $\langle {\bf s}_i{\bf s}_l
\rangle_{1,2}=-1/4$) is present up to $\alpha_1^{c}=1/4$ for both the LC
and the LM case. At this point a second-order phase transition leads to
the non-saturated magnetic structure. For the case of two coupled LM
systems we have that $\langle {\bf s}_i{\bf s}_j
\rangle_{1,1}-1/4\sim(\alpha_1-1/4)$ while for the case of two coupled
LC's $\langle {\bf s}_i{\bf s}_j
\rangle_{1,1}-1/4\sim\sqrt{\alpha_1-1/4}$, both for small
$(\alpha_1-1/4)$. The latter result can be rigorously established from
the analytic expression
$E_{\gamma}^{LC}(x_{\gamma})-1/4=-(1-x_{\gamma})+(\pi^2/48)
(1-x_{\gamma})^3+O((1-x_{\gamma})^5)$ for the energy of the
antiferromagnetic LC near maximal polarization, $x_{\gamma}$, which can
be extracted from the Bethe-Ansatz \cite{bethe}.

The first- and second-order phase boundaries shown in Fig.1 meet in
tricritical points at $(\alpha_1,\alpha_2)=1/4\hspace{3pt}(1 \ , \ -1 +
{1 \over \ln 2})$ and at $(\alpha_1,\alpha_2)=1/4\hspace{3pt}(-1 + { 1
\over \ln 2} \ , \ 1)$ for the LC case and in a tetracritical point at
$1/4\hspace{3pt}(1,1)$ for the LM case. The line of the first order
transition is the result of a numerical solution of the Bethe-Ansatz for
the LC system, whereas for two coupled LM systems it is given by $16
\alpha_1 \alpha_2 = 1$. Finally we note that the straight first-order
line between the two tricritical points in Fig.1(a) separates the state
with classical ferromagnetic long-range order within the subsystems
($x_{\gamma}=1, \gamma= 1,2$) from the quantum disordered state without
long-range order (Bethe singlet).


{\it Thermodynamics}. --
For the Lieb-Mattis model the saddle-point approximation becomes exact,
i.e. the partition function is given in the thermodynamic limit by its
largest term \cite{baxter}. The same statement holds also for the model
of coupled Lieb-Mattis subsystems considered here and all thermodynamic
potentials can be determined. The quantum numbers of the largest term in
the partition function depend on both $\alpha_1$, $\alpha_2$ and the
temperature $T$. For the Lieb-Mattis model the four (normalized)
quantum numbers $x^{A,B}_{\gamma}$ of the total spin operators $({\bf
S}^{A,B}_{\gamma})^2$ of the respective sublattices are independently
conserved. Due to symmetry we have $x^{A}_{\gamma}=x^{B}_{\gamma}\equiv
y_{\gamma}$ ($\gamma=1,2$) and the total subsystem magnetisation
$x_{\gamma}\in [0,y_{\gamma}]$. The subsystem magnetisations
$x_{\gamma}=2 S_{\gamma}/N_{\gamma}$ are always antiparallel and
$S=|S_2-S_1|$. We can define three independent order parameters
$\psi_1,\ \psi_2$ and $\psi_3$ by
\begin{equation}\label{order_parameter}
\psi_1 = y_1^2-x_1^2 \hspace{0.15cm};\hspace{0.15cm}
\psi_2 = y_2^2-x_2^2 \hspace{0.15cm};\hspace{0.15cm}
\psi_3 = x_1^2+x_2^2 \hspace{0.15cm}.
\end{equation}
The parameters $\psi_1$ and $\psi_2$  are proportional to the
antiferromagnetic order parameter, \linebreak $\psi_{\gamma} =
4(1/N_{\gamma}^2) \langle ({\bf S}_{\gamma}^A - {\bf S}_{\gamma}^B)^2
\rangle $, of the respective subsystems ($\gamma = 1,2$) and $\psi_3$ is
proportional to the order parameter for ferromagnetism in the individual
subsystems, \linebreak $\psi_3 = 4 \big [ (1/N_{1}^2) \langle {\bf
S}_{1}^2 \rangle + (1/N_{2}^2) \langle {\bf S}_{2}^2 \rangle \big ]$ .
Though there are $2^3=8$ possible phases, one of them, $\psi_i>0,\
i=1,2,3$, is found to be thermodynamically never stable. In the table we
list the stable phases and illustrate in Fig.3 the phase diagram as a
function of $\alpha_1$ and temperature $T$ for a constant
$\alpha_2=0.075$. The high-temperature paramagnetic P-phase has the
highest symmetry. For small $\alpha_{\gamma}$ ($\gamma=1,2)$ the
inter-sublattice exchange dominates and  both the first and the second
subsystem are ferromagnetic. We call the resulting phase the
ferromagnetic, or F-phase. For large $\alpha_{\gamma}$ ($\gamma=1,2)$ the
intra-sublattice couplings dominate, leading to a singlet state in both
subsystems, corresponding to the collinear antiferromagnetic AF-phase.
The low temperature AF-phase gives way to the AF$_{1}$-phase at higher
temperature when the second subsystem becomes paramagnetic, $x_2=0$ (the
AF$_2$ phase is realized for certain $\alpha_1<\alpha_2$).

The symmetry group of the noncollinear {\it ``order-from-disorder''}
OD$_1$-phase (the OD$_2$-phase is realized for certain
$\alpha_1<\alpha_2$) is a subgroup of the symmetry groups of both the
F-phase and AF$_1$-phase (compare the table), which in turn are
subgroups of the fully symmetric P-phase. These relations are consistent
with general requirements for the tetracritical point given by
$\alpha_1=\sqrt{(1/4)\alpha_2^2 + 1/8} -(1/2)\alpha_2$ and $ k_B T=
(1/2)\alpha_1$, in which these four phases meet \cite{landau}. All phase
transitions shown in Fig.3 are of second order (with the characteristic
jump in the specific heat), besides the transition from the OD$_1$-phase
to the AF-phase. The symmetry group of neither the OD$_1$-phase nor of
the AF-phase is a subgroup of the other (see the table) and the
transition is consequently of first-order.

For every region of the phase diagram a distinct set of self-consistency
equations determines the free energy and the correlation functions.
%
%
%
%
E.g., the
subsystem magnetisations $x_1=x_2/(4\alpha_1)$ and $x_2\equiv y_2$ as
well as the sublattice magnetisations of the first subsystem, $y_1$, in
this OD$_1$-phase are
determined via the equations
\begin{equation}\label{self_consistency}
y_1 = \tanh\left[ \, \alpha_1 y_1/(2 k_B T)\, \right]
\hspace{0.5cm};\hspace{0.5cm}
x_2 = \tanh\left[ \, (1/\alpha_1-8\alpha_2) \, x_2/(16 k_B T)\, \right].
\end{equation}
The inter- and intra-subsystem correlation functions for the
OD$_1$-phase are given by
\begin{equation}\label{correlations}
\langle {\bf s}_i{\bf s}_l\rangle _{1,2} \hspace{-0.15cm} =
\hspace{-0.15cm}
    - x_2^2\, /\, (16\alpha_1) \hspace{0.15cm} ; \hspace{0.15cm}
    \langle {\bf s}_l{\bf s}_m\rangle _{2,2} =
     x_2^2\, /\, 4  \hspace{0.15cm} ; \hspace{0.15cm}
\langle {\bf s}_i{\bf s}_j\rangle _{1,1} \hspace{-0.15cm} =
\hspace{-0.15cm}
     x_2^2\, /\, (32\alpha_1^2)
   -  y_1^2\, /\, 4 \hspace{0.2cm},
\end{equation}
where $i,l\in A$ and $j,m\in B$ in $\langle {\bf s}_i{\bf s}_j\rangle
_{1,1}$ and $\langle {\bf s}_l{\bf s}_m\rangle _{2,2}$. The transition
from the OD$_1$-phase to the AF$_1$-phase takes place when $x_2=0$ in
Eq.\ (\ref{self_consistency}) and the transition to the F-phase takes
place when $y_1=x_1=x_2/(4\alpha_1)$.

In the OD$_1$-phase the sublattice magnetisations in the first subsystem
are neither parallel nor anti-parallel, one might call the OD$_1$-phase
also a {\it ``twist''}-phase. The temperature dependence of the
inter-sublattice correlation functions in the first subsystem, $\langle
{\bf s}_i{\bf s}_j\rangle _{1,1}$, are presented in Fig.4 for some
selected values of $\alpha_1$ and $\alpha_2=0.075$. In the OD$_1$-phase
($1/4<\alpha_1<1/(16\alpha_2)$ at $T=0$) the correlation function
$\langle {\bf s}_i{\bf s}_j\rangle _{1,1}$ increases in magnitude as a
function of temperature. This effect is particulary pronounced for
$\alpha_1=1/\sqrt{8}$. For $\alpha_1=1/\sqrt{8}$ the A- and the B-
sublattices of the first subsystem are completely uncorrelated at zero
temperature (see Eq.\ (\ref{correlations})) and become more and more
correlated with increasing temperature, as long as we remain in the
OD$_1$-phase. This phenomenon is called {\it order from disorder}
\cite{villain,shender82}, and is due to the interplay between
different competing
energy scales govering the correlations (compare Eq.
(\ref{self_consistency}) and Eq. (\ref{correlations})) of the individual
subsystems. The disappearance of $\langle{\bf s}_i{\bf
s}_j\rangle _{1,1}$ at zero temperature for $\alpha_1=1/\sqrt{8}$ and
the increase of magnetic order by fluctuations is
related to the maximum in the GS energy $E_0(\alpha_1)$ which
signals maximal frustration.
On the other hand, the
correlations decrease in the usual way with increasing temperature
within the ferromagnetic phase or the antiferromagnetic phase.

{\it Summary.} --
We have discussed quantum spin Heisenberg systems built of two coupled
subsystems. If the individual subsystems are integrable on their part,
the whole system is solvable for a long-ranged inter-subsystem coupling
of constant magnitude. The three antiferromagnetic exchange integrals,
namely the two intra-subsystem couplings and the inter-subsystem
exchange parameter compete with each other. This competition gives rise
to interesting frustration effects. Varying the strength of the three
couplings we can tune the magnetic ordering of the individual subsystems
between ferromagnetic and antiferromagnetic. In the limit of maximal
competition between different exchange couplings neither
ferromagnetic nor antiferromagnetic correlations
dominate in the ground state.
However,  additional thermal fluctuations change the interplay
between the competing energy scales and lead to an increase of magnetic order
with temperature.
This mechanism driving the {\it order from disorder} scenario seems
to be not restricted on the special Lieb-Mattis coupling and we argue
that the same phenomenon should be present in systems with more
realistic couplings.


We are indebted to  N.B. Ivanov for fruitful discussions.
This work was supported by the DFG (Ri 615/1-1).


\newpage


\begin{table}
\caption{
The seven thermodynamically stable phases for two coupled Lieb-Mattis
antiferromagnets. $y_{\gamma}= (4/N_{\gamma})S_{\gamma}^{A,B}$ ($0 \le
y_{\gamma} \le 1,\ \gamma=1,2$) are the relative magnetisations of the
A,B-sublattice of the first/second subsystem and $x_{\gamma}=
(2/N_{\gamma})S_{\gamma}$ ($0 \le y_{\gamma} \le 1$)
$x_{\gamma}\in[0,y_{\gamma}]$ are the magnetisations of the respective
($\gamma=1,2$) subsystem. The paramagnetic (P), ferromagnetic (F), the
three antiferromagnetic phases (AF$_1$), (AF$_2$), (AF) and the {\it
order from disorder} phases (OD$_1$) and (OD$_2$) are characterized by
the order parameters $\psi_1 = y_1^2-x_1^2$, $\psi_2 = y_2^2-x_2^2$ and
$\psi_3 = x_1^2+x_2^2$.}
\vspace{0.5cm}
\begin{tabular}{l|cc|cc|ccc}
      & $y_1$ & $y_2$ &  $x_1$ &   $x_2$ & $\psi_1$ & $\psi_2$ & $\psi_3$
  \\ \hline\hline
P     &     0 &     0 &      0 &       0 &        0 &        0 &       0
  \\ \hline
F     &     + &     + & $y_1$ & $y_2$ &        0 &        0 &       +
  \\ \hline
AF$_1$&     + &     0 &      0 &       0 &        + &        0 &       0
  \\
AF$_2$&     0 &     + &      0 &       0 &        0 &        + &       0
  \\
AF    &     + &     + &      0 &       0 &        + &        + &       0
  \\ \hline
OD$_1$&     + &     + & $\frac{y_2}{4\alpha_1}$
                               & $y_2$ &        + &        0 &       +
  \\
OD$_2$&     + &     + & $y_1$ &
                $\frac{y_1}{4\alpha_2}$ &        0 &        + &       +
  \\
\end{tabular}
\end{table}


\begin{figure}
\caption{The ground-state phase diagram of a) two coupled
              antiferromagnetic linear chains and b) two coupled
              Lieb-Mattis antiferromagnets, as a function of the
              intra-subsystem couplings $\alpha_1$ and $\alpha_2$.
              The inter-subsystem coupling is $\alpha=1$. The phases are
              characterized by the
              magnetisations $x_1$ and $x_2$ of the
              subsystems. The dashed/full lines denote
              phase transitions of first/second order.}
\end{figure}

\begin{figure}
\caption{
The ground-state inter-subsystem
              $\langle {\bf s}_i{\bf s}_l\rangle _{1,2}$ and the
              intra-subsystem $\langle {\bf s}_i{\bf s}_j\rangle _{1,1}$
              correlation function for two coupled linear chains
              (LC, dashed lines) and two coupled Lieb-Mattis
              antiferromagnets (LM, full lines), as a function
              of $\alpha_1$.
              The intra-subsystem coupling $\alpha_2=0.075$
              and the inter-subsystem coupling $\alpha=1$. At
              $\alpha_1=1/4$ a phase transition of second-order
              occurs for both systems
              and the correlation functions have a square-root
              singularity (kink) for the case of two LC's (LM's).
}
\end{figure}
\begin{figure}
\caption{
Phase diagram of two coupled Lieb-Mattis antiferromagnets,
              as a function of temperature $T$ and the intra-subsystem
              coupling $\alpha_1$, for $\alpha_2=0.075$ and the
              inter-subsystem coupling $\alpha=1$. The dashed/full lines
              denote phase transitions of first/second order. The
              phases are labelled with respect to the values of the
              correlation functions within the first subsystem.
}
\end{figure}

\begin{figure}
\caption{
The intra-subsystem correlation function
              $\langle {\bf s}_i{\bf s}_j\rangle _{1,1}$ ($i\in A,j\in
              B$)
              for two coupled Lieb-Mattis antiferromagnets, as a function
              of temperature $T$ for some selected $\alpha_1$'s, and
              fixed
              $\alpha_2=0.075$, $\alpha=1$. The kinks in
              $\langle {\bf s}_i{\bf s}_j\rangle _{1,1}$ correspond to
              phase transitions of second order. Note the
              particular strong
              {\it ``order from disorder''} phenomenon
              for  $\alpha_1=8^{-1/2}$.
}
\end{figure}
\end{document}